**Rapid synthesis of uniformly small nickel nanoparticles for the surface functionalization of epitaxial graphene**

*Ylea Vlamidis*[*], *Stiven Forti, Antonio Rossi, Carmela Marinelli, Camilla Coletti, Stefan Heun, Stefano Veronesi*[*]

Y. Vlamidis, S. Heun, S. Veronesi
NEST, Istituto Nanoscienze–CNR and Scuola Normale Superiore, Piazza San Silvestro 12, 56127 Pisa, Italy

Y. Vlamidis, C. Marinelli,
Department of Physical Science, Earth, and Environment, University of Siena, Via Roma 56, 53100 Siena, Italy
E-mail: stefano.veronesi@nano.cnr.it; ylea.vlamidis@unisi.it

S. Forti, A. Rossi, C. Coletti
Center for Nanotechnology Innovation@NEST, Istituto Italiano di Tecnologia, Piazza San Silvestro 12, Pisa 56127, Italy
Graphene Laboratories, Istituto Italiano di Tecnologia, Genova 16163, Italy

**Keywords:** Nickel nanoparticles, cetyltrimethylammonium bromide (CTAB), epitaxial graphene, X-ray photoelectron spectroscopy, scanning tunneling microscopy

**Abstract**

Nickel nanoparticles (Ni NPs), thanks to their peculiar properties, are interesting materials for many applications including catalysis, hydrogen storage, and sensors. In this work, Ni NPs are synthesized in aqueous solution by a simple and rapid procedure with cetyltrimethylammonium bromide (CTAB) as a capping agent, and are extensively characterized by dynamic light scattering (DLS), scanning electron microscopy (SEM), and atomic force microscopy (AFM). We investigated their shape, dimension, and their distribution on the surfaces of $SiO_2$ and epitaxial graphene (EG) samples. Ni NPs have an average diameter of ~11 nm, with a narrow size dispersion, and their arrangement on the surface is strongly dependent on the substrate. EG samples functionalized with Ni NPs are further characterized by X-ray photoelectron spectroscopy (XPS), as made and after thermal annealing above 350°C to confirm the degradation of CTAB and the presence of metallic Ni(0). Moreover, high resolution scanning tunneling microscopy (STM) topographies reveal the structural stability of the NPs up to 550 °C.

**1. Introduction**

In the past two decades, the synthesis of metal nanoparticles has received considerable attention thanks to their interesting and unusual size-dependent properties and potential applications. In particular, magnetic nanomaterials such as Ni NPs have attracted much attention for a variety of applications, such as sensors,[1] catalysis,[2,3] imaging,[4] fluids applications,[3] spintronic devices,[5] and energy storage.[6,7]

Currently one of the most interesting applications of Ni NPs is in the energy/hydrogen storage field. Due to its environmental friendliness and remarkable energy efficiency, hydrogen-based fuel emerges as a promising alternative for the gradual substitution of conventional energy sources. Nevertheless, effectively and safely storing hydrogen in a compact manner for practical application poses significant technical challenges. Nanostructured materials have recently gathered considerable interest as a potential medium for hydrogen storage.[8–11] In particular, interconnected graphene networks boast remarkable features such as lightweight construction, high specific surface area, significant chemical stability, and excellent on-board reversibility.[12] Aside from graphene, certain metal catalysts have the capability to effectively capture hydrogen. Hydrogen molecules attach to carbon with relatively weak van der Waals forces,[13] limiting their uptake. Consequently, current research is directed towards enhancing hydrogen absorption by decorating the graphene surface with metals, aiming to significantly increase the hydrogen uptake.[14,15] Among the transition metals, Ni atoms are particularly intriguing due to their peculiar interactions with the graphene lattice and capacity to enhance the stability of the sorbent structure.[5] These particles exhibit the capability to adsorb numerous hydrogen molecules on their surface, thereby catalyzing their dissociation into atomic hydrogen. This process aids in improving the hydrogen absorption/desorption performance in nanocomposites while also serving as anchoring sites for hydrogen dissociation.[16,17] Furthermore, nickel offers the additional advantage of being inexpensive compared to other metals, such as noble metals. The combination of graphene and Ni could greatly enhance the hydrogen storage capacity when compared to the capacity in the individual components, as already reported for carbon-based nanomaterials.[16,18]

The synthesis of transition metal nanoparticles is relatively difficult because they are easily oxidized. So far, a number of chemical routes have been investigated to synthesize Ni NPs including micro-emulsion,[19] solvothermal reduction,[3] thermal decomposition of organic complexes,[20,21] microwave,[22] sonochemical,[23] and the polyol process, using either hydrazine or sodium borohydride as reducing agent.[24,25] The polyol process restricts particle size due to the presence of glycol functional groups and also results in pure Ni without using any inert atmosphere during synthesis.[26] However, these processes can require a great deal of energy over a prolonged period of time, and the use of organic solvents.[27]

Despite the many advantages of polyol processing, as low concentration precursors, the use of hydrazine and prolonged stirring are not encouraging factors for the scale-up of the process.

In this work, we describe a direct and fast approach to fabricate monodisperse Ni NPs of about 10nm in size from nickel acetate by the reduction of nickel with sodium borohydride. The reaction occurs in mild conditions and short times, in aqueous solution containing only cetyltrimethylammonium bromide (CTAB). This surfactant is employed to control the size of the NPs during the synthesis and acts as stabilizer afterwards.[19,28] The particle size and structure of the resultant nanoparticles have been characterized by scanning electron microscopy (SEM) and atomic force microscopy (AFM). Furthermore, the synthesized NPs were employed for the functionalization of epitaxial graphene grown on 6H-SiC, and the chemical interaction between graphene and nickel as well as the physical and spectroscopic properties were investigated by X-ray photoelectron spectroscopy (XPS) in the as-prepared samples and after annealing at temperatures higher than that typical of CTAB decomposition.[29,30] Finally, scanning tunneling microscopy (STM) and spectroscopy (STS) provided local details regarding the surface configuration and electronic properties, giving additional insight in the interactions between graphene and Ni atoms, along with the stability of Ni after thermal annealing up to 550°C.

The comprehensive characterization of Ni NPs and the EG/Ni system here described provides valuable insight into the applications of the proposed synthesis and functionalization procedure.

## 2. Results and Discussion

### 2.1. Ni NPs characterization

In this work, we synthesized and characterized Ni NPs/CTAB to investigate, respectively, the size, shape, and eventually their interaction with epitaxially grown graphene. Ni NPs were synthesized following the protocol described in the Experimental section. Briefly, Ni NPs were synthesized by fast reduction of an aqueous solution of nickel acetate by sodium borohydride at about 1 °C in the presence of CTAB. A full set of characterizations was performed to check the products and the reproducibility of the protocol. The synthesis of Ni nanoparticles involves two primary steps: initially, Ni ions undergo reduction to form metal atoms, followed by nucleation of these reduced Ni atoms. Subsequently, the nickel atoms aggregate in clusters of growing dimensions, leading to enhanced stability of the nanoparticles. Finally, stabilizing agents cap the metal atom clusters to yield stable Ni NPs. These steps are influenced by the reaction system and parameters controlling the reaction. During the synthesis process, the growth of particles is affected by the concentration of reactants and the synthesis temperature. Generally, higher temperatures and concentrations result in a faster rate of coagulation, providing more opportunities for nuclei to grow into larger particles. In our experiment, we conducted the synthesis at low temperature to slow down the reaction kinetics and inhibit the growth of large Ni nuclei.

To investigate the interaction between the Ni NPs and the capping agent, Fourier transform infrared (FT-IR) analysis was carried out. The vibrational spectra, shown in **Figure 1a**, were normalized for an easier comparison. The high noise level in the FT-IR spectra of Ni NPs/CTAB is attributed to the lower signal intensity. For pure CTAB, the peaks at 2918 and 2848 $cm^{-1}$ are attributed to C-H stretching vibration of methyl and methylene groups.[31] The symmetric and asymmetric stretching vibrations of the ($CH_3-N^+$) group appear between 1440 and 1500 $cm^{-1}$.[31] The band at 960 $cm^{-1}$ corresponds to the out-of-plane C-H vibration of $CH_3$, while the band at 720 $cm^{-1}$ is assigned to $Br^-$.[31]

Compared to the pure CTAB spectrum, there is a variation in the characteristic absorption bands of the Ni NPs/CTAB samples prepared (refer to **Figure 1a**). Indeed, the characteristic absorption bands of CTAB in the nickel sample appeared shifted to 1500 $cm^{-1}$ and 1715 $cm^{-1}$ ($v(CH_3-N^+)$). The C-H stretching vibration appear in the range between 2830-2950 $cm^{-1}$. The broad bands between 3200 and 4000 $cm^{-1}$ are due to O-H stretching in water and ethanol.[32] This demonstrates that mutual interactions between the ammonium moiety in CTAB and the surface of the Ni NPs have taken place.[2]

When the surfactant CTAB is dispersed in deionized water, a certain critical micelle concentration is formed. This surfactant can induce nickel nanocrystalline growth with different orientation which results in different shaped nickel nanoparticles.[33] The nucleation and crystal growth of Ni NPs are supposed to proceed in the inner shells of the micelles. As demonstrated by FTIR analysis, the inner chains of the micelles were composed of nitrogen-containing bonds ($CH_3-N^+$ group). The shape and stabilization of the resulting nickel nanoparticles may be due to the coordinative interaction between the nitrogen-containing bonds in CTAB and the nanostructured nickel.

Dynamic light scattering (DLS) is an important tool for determining the size of synthesized Ni NPs/CTAB in solution, as well as for predicting their long-term stability. The DLS analysis revealed that the synthesized Ni NPs show an average hydrodynamic diameter of 22.3±1.7 nm (**Figures 1b and c**). The polydispersity index (PDI) of 0.34 indicates a quite uniform size distribution and a good stability of the suspension over time, with low aggregation tendency.[34] The hydrodynamic diameter provides information about the inorganic core along with any coating material and the solvent layer attached to the particle as it moves under the influence of Brownian motion. The radius is calculated from the diffusion coefficient in a given solvent. Nevertheless, the hydrodynamic diameter of a nanoparticle in ethanol obtained by the DLS method is generally overestimated if compared to that directly measured from SEM or TEM images, due to the hydration of the nanoparticles and the low density of the solvent.[35]

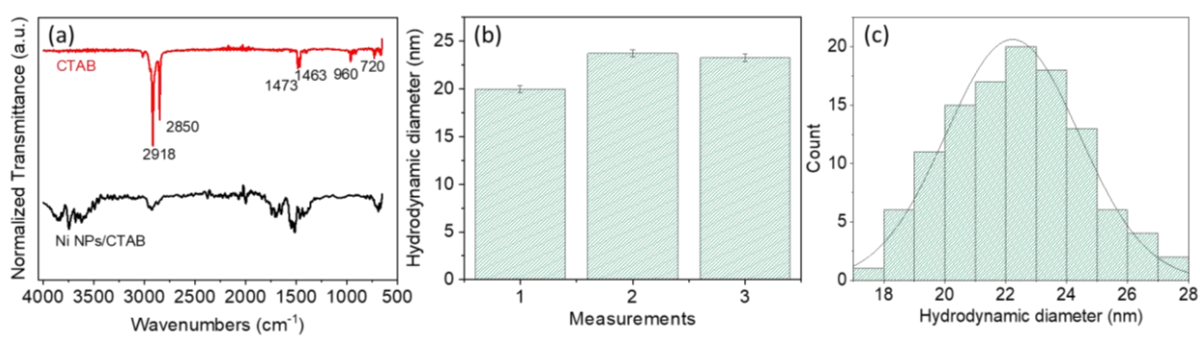

**Figure 1.** (a) FT-IR spectra of pure CTAB and Ni NPs/CTAB, DLS measurements for (b) hydrodynamic diameter and (c) related size distribution of Ni NPs/CTAB in ethanol/water, resulting from three independent measurements.

Initially, to study the morphology of the synthesized materials, the Ni NPs were deposited on Si/SiO$_2$ substrates. **Figure 2** shows the AFM topography and SEM morphology of Ni NPs obtained via chemical reduction. As illustrated, the reaction carried out at low temperature promotes controlled nucleation and growth, leading to the formation of small and quite uniform nanoparticles. From the AFM height image, several aggregates can be seen (**Figure 2a**), whilst the phase image (**Figure 2b**) reveals the presence of smaller individual NPs, besides larger aggregates. Ni NPs display a spherical shape with diameters ranging from 5 to 15 nm, determined from the AFM profiles (inset of **Figure 2b**). The average diameter is 11.2±2.9 nm, estimated from the size distribution histogram obtained from several SEM images, shown in **Figure 2d**. SEM allows for the imaging of bare metallic nanoparticles, thanks to the clear topographic contrast between metal surface and the organic shell, therefore the estimated dimensions are more reliable compared to those obtained from DLS for assessing the effective diameter of the Ni NPs. A certain extent of aggregation of the particles upon solvent evaporation can be observed which might result mostly from magnetic interactions between the particles.[27]

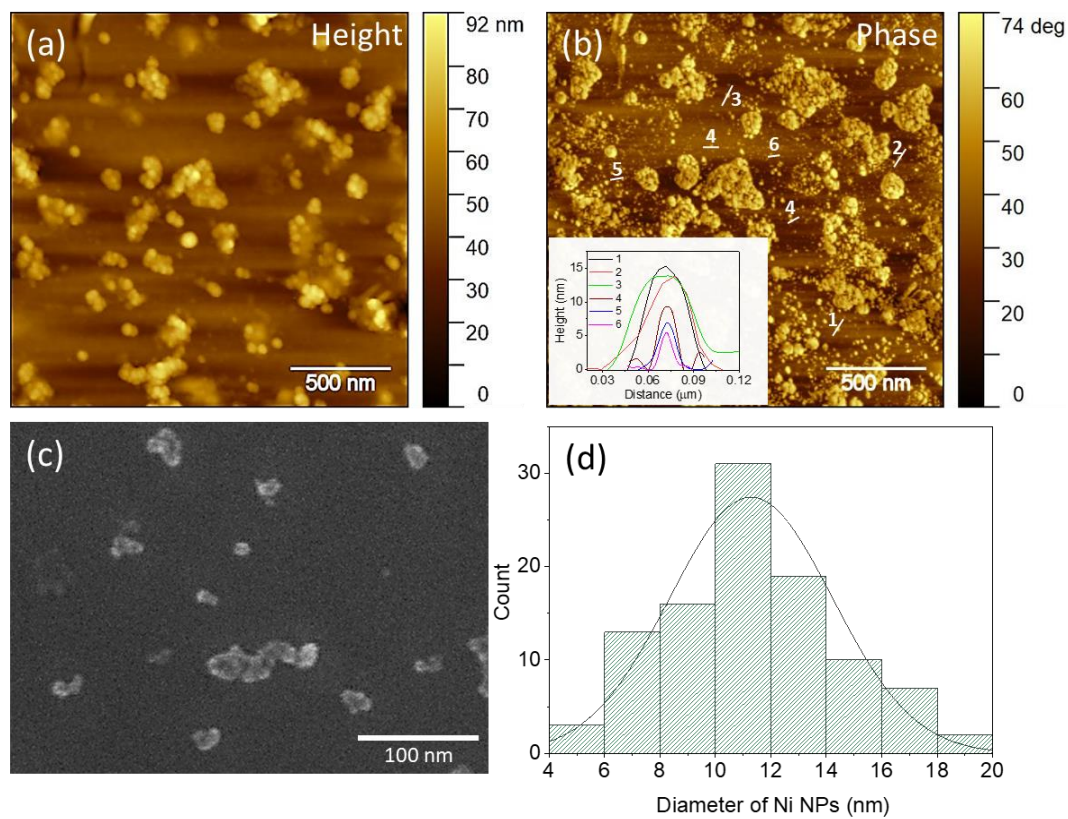

**Figure 2.** Morphological characterization of Ni NPs on Si/SiO$_2$. (a) AFM height and (b) phase images with profile distribution in the inset. (c) SEM image and (d) particle size histogram derived from observations of more than 100 Ni NPs using SEM.

**2.2. Morphological and spectroscopic characterization of Ni NPs on epitaxial graphene**

Further, we investigated the interactions and oxidation state of Ni NPs dispersed on epitaxial graphene both as deposited on EG and after performing annealing steps at different temperatures (350 and 550°C) to monitor possible changes in the binding states of the elements at temperatures above the CTAB decomposition temperature.[30]

*2.2.1 Ni NPS on epitaxial graphene (sample EG1)*

EG samples were grown on 6H-SiC(0001) as outlined in the Experimental section. Firstly, Raman spectroscopy was exploited to assess the quality and composition of EG. Two different EG samples have been utilized, one composed of mono/bilayer, and the other composed of buffer layer/monolayer, herein referred to as EG1 and EG2, respectively. EG1 was employed for the XPS analysis, while EG2 for the STM investigation. The composition of the two EG samples was confirmed by Raman spectroscopy, reported in the S.I.

*2.2.2. Morphological characterization of Ni NPs on EG1*

The surface morphology and distribution of Ni NPs on epitaxial graphene were characterized by SEM and AFM images. AFM micrographs of the surface of EG1 decorated with Ni NPs, at different magnification, are shown in **Figures 3a** and **3b**. The topographies reveal the presence of atomic terraces, of width of hundreds of nm to micrometers, typically observed after the growth of graphene with good crystallinity and thickness homogeneity. Ni NPs appear to cover the entire surface. However, they reveal a tendency to arrange preferentially in bigger clusters due to a coalescence phenomenon and higher stability along the step edges of the surface. This phenomenon was already observed also for other metal nanoparticles and represents the energetically most stable configuration.[5,36–38] SEM images (**Figure 3c** and **3d**) display a smooth surface where graphene domains decorated with Ni NPs can be observed.

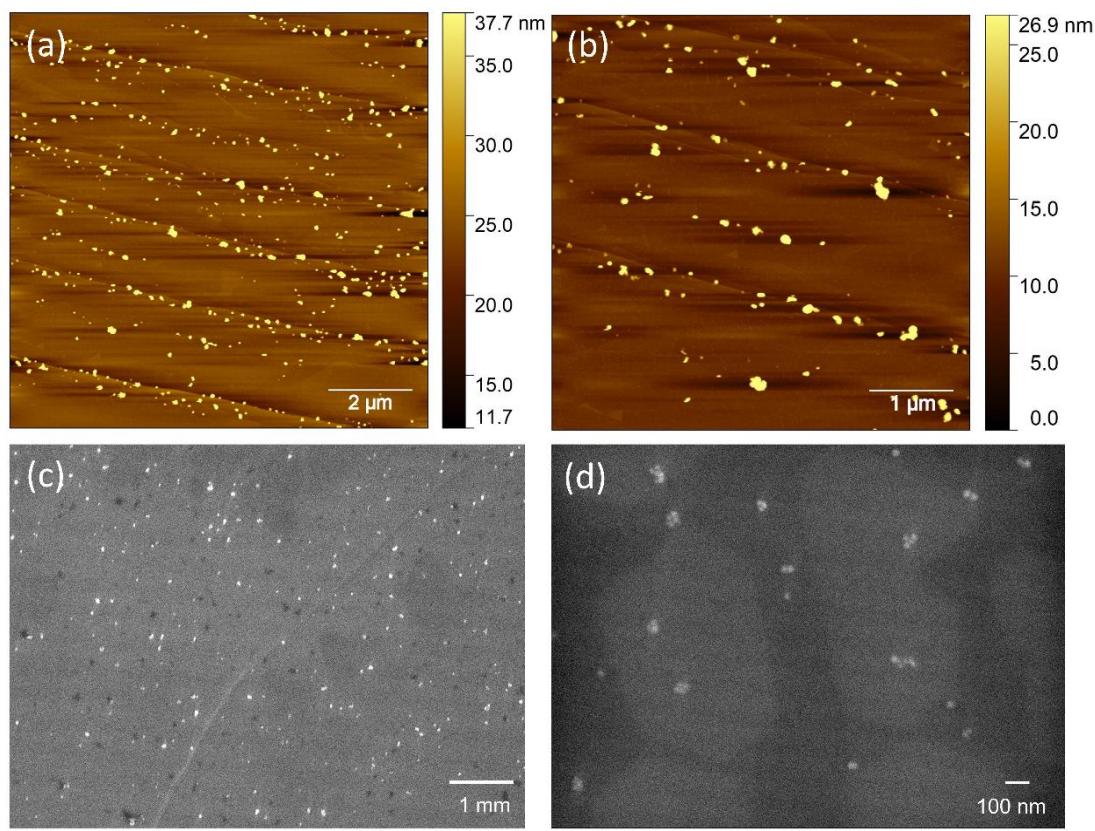

**Figure 3.** Characterization of Ni NPs on epitaxial graphene (sample EG1): (a) AFM topography and (b) magnified view of the area showing that the steps are clearly decorated by NPs clusters. (c and d) SEM images at different magnifications. In panel (d) the brighter areas indicate the presence of randomly stacked bilayer islands.

*2.2.3. XPS analysis of Ni NPs on EG1*

AFM topographies reveal that the morphology of sample EG1 decorated with Ni NPs is not significantly affected by the thermal annealing at temperatures up to 550°C (see **Figure S2**). XPS measurements were carried out to investigate the effect of the annealing process at 350°C and 550°C on the chemical state and composition of the sample surface. **Figure 4** reports the spectra acquired at room temperature (prior to thermal treatment) and after the two annealing steps for Ni 2p, C1s, and Si 2p core levels. From **Figure 4a** we can notice how the Ni 2p spectrum at room temperature displays three main doublets: one with the high-spin component at 852.8 eV (Ni $2p_{3/2}$) is assigned to metallic $Ni^0$, one at 855.8 eV, ascribed to oxides and one at 863.0 eV, which we assign to hydroxides.[39-41] Upon annealing at 350°C and 550°C, we observe a remarkable decrease in intensity of the oxide and hydroxide related components, in favor of the low oxidation state component, which becomes dominant, implying a conversion of the nickel content towards a metallic state. This observation is consistent with the occurrence of thermal degradation of CTAB micelles, as well as with the evaporation of the solvent from the surface at the explored temperatures.[29] No significant differences in the relative ration of the components are notable comparing the spectra acquired after annealing at 350 °C and 550 °C, suggesting that the surfactant decomposes mostly below 350°C.

The temperature evolution of the C 1s core level spectra is shown in **Figure 4b**. In the deconvoluted spectra, the component located at∼283.8 eV (red line) is assigned to SiC. The peaks at 284.9 eV and 285.6 eV are attributed to the S1 and S2 components (dark and light grey, respectively), which describe the interaction of the interfacial buffer layer with the substrate according to the literature.[42] Graphene is modelled with an asymmetric lineshape (Doniach-Sunjic) centered at 284.7 eV and its intensity is consistent with the presence of a bilayer (see also Ref. [43]). The CTAB contains $CH_2$ groups which are therefore described by a symmetric Voigt function centered at 285.6 eV (yellow line), which indeed disappears when the sample is heated in UHV at temperatures above 350 °C. In turn, above that temperature we observe the emergence of another component as lower binding energy, which we ascribe to a small fraction of the nickel interacting with graphene via the formation of carbides $Ni_3C$. This (green line) component is located at 283.6 eV.

In panel c we show the thermal evolution of the Si 2p core level. The spectrum recorded at room temperature can be fitted with two components: a main doublet at 101.5 eV, associated to the signal coming from the bulk SiC substrate (red line) and a minority component at higher binding energy (blue line), which is usually associated with defects.[42] With increasing temperature, a high-oxidation state component emerges (green line). We assign this component to the interaction of the interfacial Si atoms with the oxygen-rich groups originating from the decomposition of the solvent. We point out that there is no evidence of a low-binding energy component that could be assigned to nickel silicides, indicating the at this stage, nickel does not intercalate at the heterointerface between SiC(0001) and its carbon-rich $6\sqrt{3}$ reconstruction.

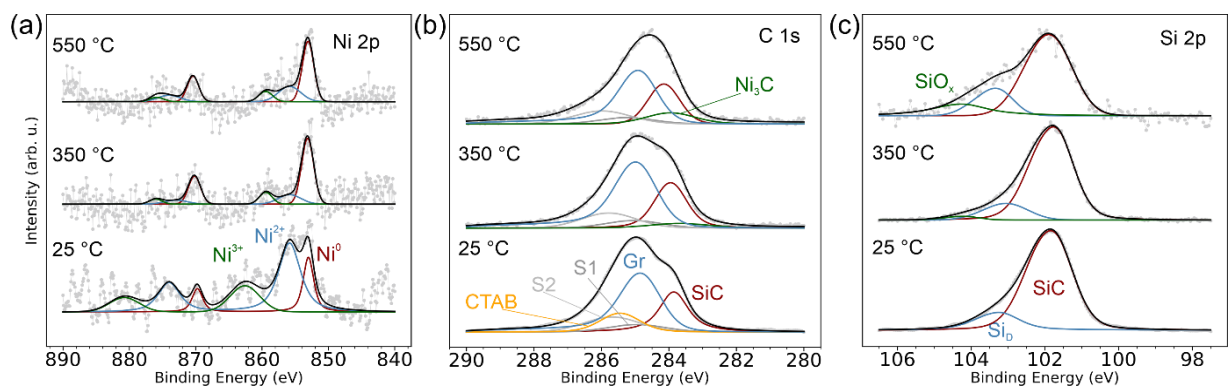

**Figure 4.** High resolution deconvoluted XPS spectra of EG grown on 6H-SiC (0001) (sample EG1) decorated with Ni NPs at room temperature and after annealing steps at 350°C and 550°C: (a) Ni 2p, (b) C 1s, and (c) Si 2p core levels. Experimental data (circles) are shown together with fitted results (line).

*2.2.4 STM characterization of epitaxial graphene decorated with Ni NPs*

In order to deeply investigate the surface morphology with atomic resolution, a sample of graphene constituted of buffer layer with a low coverage of monolayer (sample EG2) was grown and Ni NPs were deposited on the surface with the same method described before.
The composition of the EG2 sample was confirmed by Raman spectroscopy and AFM topography (refer to **Figures S3 and S4**).
STM allowed for the in-depth characterization of the surface topography. **Figure 5** reports four STM representative scans, at different magnifications, of the epitaxial graphene surface decorated with Ni NPs upon annealing in ultra-high vacuum (UHV) at 550 °C. In particular, the images were acquired in monolayer regions, to better distinguish the NPs on the top of the surface. From STM imaging, the presence of clusters of Ni NPs along the step edges, and on the terraces can be clearly detected as bright protrusions. The NPs are arranged in clusters as large as 20-30 nm, with a height of about 2-3 nm, whereas the average diameter of the NPs is about 5 nm, as estimated by a statistical analysis based on the STM images. Accordingly to those estimations the nanoclusters appear flattened, and this effect is reasonably due to the elevated temperature and the strong interaction between Ni and C atoms. Both the coalescence and the arrangement along the step edges after the thermal annealing demonstrate a high mobility of Ni NPs at the explored temperature.[5,37] Higher magnifications (**Figures 5d and e**) show both isolated NPs or small aggregates and the typical quasi (6x6) periodicity with a corrugation of ~1.84 nm observed on epitaxial graphene [44,45] (see profile reported in **Figure 5f**). The modulation with a 0.249 nm period showed in the tunneling current profile (**Figure 5f**) is assigned to the graphene honeycomb lattice.[46] From the STM images, individual NPs can be resolved, and analyzing a few Ni NPs, the dimensions result rather uniform and on average ranges between 3 and 6 nm (refer to STM profiles shown in **Figure S5**).

Scanning tunneling spectroscopy (STS) was also performed on the sample to further investigate the local electronic properties of EG2/Ni NPs. STS measurements were performed both on the graphene layer and on clusters of Ni NPs (refer to **Figure S6**). The typical current-voltage characteristics measured above several Ni NPs was rather featureless and linear in the explored bias window (between -0.2 and 0.2V), confirming a metallic behavior (**Figure S6a**). A representative STS dI/dV spectrum acquired on a flat area of monolayer graphene is shown in **Figure S6b** for comparison. The Dirac point positioned at 0 V is indicative of a neutral graphene. Since epitaxial graphene is intrinsically n-doped,[47] the observed doping level shift could be due to charge transfer between graphene and Ni.

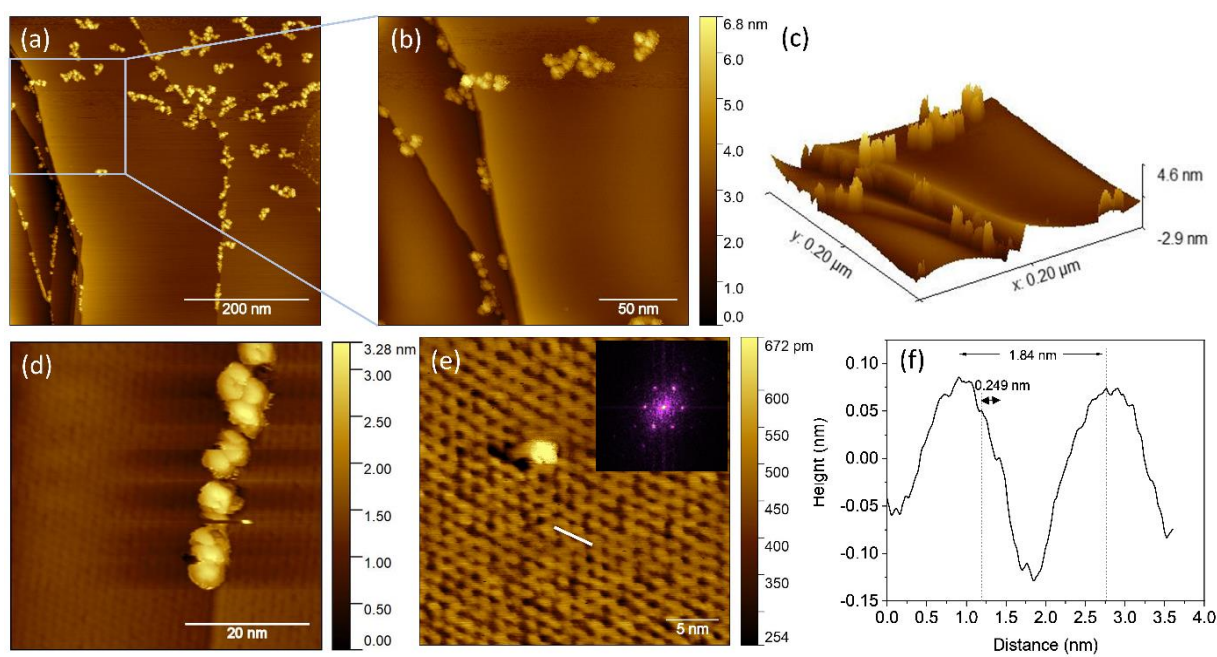

**Figure 5.** STM images of Ni NPs on sample EG2, after annealing at 550°C. (a) Image of 500x500 nm$^2$ and (b) a zoom-in (200x200 nm$^2$) into the square area indicated in (a) show small clusters of NPs on the monolayer graphene regions. Panel (c) displays a 3D visualization of (b). (d) Expanded views of 50x50 nm$^2$ and (e) 30x30 nm$^2$ acquired in different areas on the surface display the detailed morphology of the clustered or isolated Ni NPs. Inset in (e) report the 2D-FFT (2D Fast Fourier Transform) of the image. Panel (f) reports a profile obtained from the scan in panel (e), indicated by the white line. Imaging parameters: (a, b) $V_s$= 0.6 V, $I_t$ = 0.50 nA, (d, e) $V_s$= 0.4 V, $I_t$ = 0.30 nA.

## 3. Conclusions

In summary, we described a facile and fast synthesis of nickel nanoparticles achieved through the reduction of Ni(II) in a monosurfactant aqueous solution, carried out at low temperature. The synthesized Ni nanoparticles were homogenous in size, with a mean particle size around 11 nm, as obtained from AFM and SEM observation. The chemical composition of the sample surface, and in particular the decomposition of the surfactant (CTAB) upon thermal annealing up 550°C, were investigated by performing XPS analysis of Ni NPs on EG. Above 350°C we clearly observed the conversion of the nickel content towards a metallic state confirming. Furthermore, after annealing the sample at 550 °C, STM topographies revealed a preferential arrangement of the NPs in clusters located near the step edges.

The possibility to obtain small NPs and easily functionalize epitaxial graphene samples, along with the recovery of metallic Ni after the thermal degradation of CTAB at relatively low temperatures (~350°C), makes this strategy compelling for developing sensors or devices with enhanced catalytic performance. Given these interesting results, the proposed method can potentially be used also to functionalize other types of substrates with complex geometries, such as porous materials, through the diffusion of Ni nanoparticles.

## 4. Experimental Section

*Materials:* Cetyltrimethylammonium bromide (CTAB, ≥98% pure), Nickel(II) acetate tetrahydrate (98% pure), $NaBH_4$ (>98% pure) were purchased by Merck. Ultrapure water purified using a Millipore Milli-Q water system was used to prepare aqueous solutions. Ethanol was supplied by Carlo Erba Reagents. All reagents were used as received without further purification.

*Synthesis of Ni NPs:* The surfactant-assisted synthetic method is a convenient and effective pathway in the synthesis of nanoparticles, for it allows for metal nanoparticles to be precisely adjusted in terms of their size and shape.

Ni NPs with a hydrodynamic diameter of approximately 22 nm were prepared according to the following procedure. First, a solution of cetyltrimethylammonium bromide (CTAB, 12 mM) and of $Ni(OAc)_2$ (25 mM) in milliQ water (20 mL) was prepared. CTAB is a cationic surfactant acting as stabilizer for the NPs. The solution was cooled down to 1-2°C in an ice bath for 30 minutes before the reaction. While vigorously stirring, 200 μL of sodium borohydride (20 mg/mL in milliQ water) were added in 5 aliquots, and the mixture was stirred for 2 minutes more. Since after metal reduction the solution becomes dark grey, the solution is quickly collected in an Eppendorf soon after the color change and centrifuged at 13000 rpm for 10 minutes, to remove the precipitate containing large nanoparticles (diameter above ~20 nm). The smaller nanoparticles are instead suspended in the supernatant. These are collected by dilution of the solution with ethanol and washed several times with water by centrifugation and re-suspension, to remove excess reagents and/or residual impurities in the mixture. Finally, the NPs were suspended in a small volume of water (2 mL), and stored at 4°C.

*Epitaxial graphene growth:* epitaxial graphene was grown on 6H–SiC(0001) by thermal decomposition of the substrate at high-temperature in a Black Magic (Aixtron) cold wall reactor under Ar atmosphere.[48] Two different EG samples were grown, one characterized by the presence of mixture of monolayer and bilayer graphene, with bilayer coverage of at least 70% (EG1), while the other is characterized by a mixture of monolayer and buffer layer with a monolayer coverage of about 30% (EG2).

*Spectroscopic characterization of EG samples:* the composition and homogeneity of graphene was investigated by Raman spectroscopy. Raman mapping (21x21 m$^2$) was performed using a Renishaw inVia system with a 100x objective (NA 0.85), equipped with a 532 nm laser. The measurements were carried out with a1800 mm/l grating, with a fluence of about 35 mJ/μm$^2$.

*Ni NPs characterization:* Ni NPs were centrifuged and suspended in a mixture of ethanol/water (75:25) in order enable the formation of CTAB micelles and, at the same time, increase the affinity towards hydrophobic substrates such as graphene. The colloidal solution was sonicated for 10 min at room temperature to obtain a uniform dispersion. Either a Si substrate with SiO$_2$ passivation layer (250 nm) or epitaxial graphene were used as substrates in this work. The substrates were cleaned in isopropyl alcohol, dried under nitrogen flow, and then soaked into the colloidal solution (0.05 mg mL$^{-1}$) for several minutes. Then the samples were allowed to dry.

Ni NPs samples were analyzed with respect to size, shape, and dispersion using atomic force microscopy (AFM, Bruker Dimension Icon) operating in tapping mode, with 1 Hz scan rate and 512 scan lines, and in a scanning electron microscopy.

Scanning electron microscopy (SEM) images were acquired with a Jeol JSM-7500F instrument operated at an acceleration voltage of 20 kV. SEM and AFM images were employed to assess the distribution and density of Ni NPs after the functionalization of epitaxial graphene, besides their average dimension. The size distribution of Ni NPs was determined from SEM images using the software ImageJ (Washington, DC, USA).

Measurements by DLS were performed at 25 °C in a 1-mL polypropylene cuvette on a Zetasizer nano-ZS DLS (Malvern Instruments, Malvern, United Kingdom) following the manufacturer's instructions. Aqueous colloidal solutions were analyzed with a single scattering angle of 90°. Each value reported is the average of five consecutive measurements.

The vibrational spectra (FT-IR) of samples were recorded with a Perkin–Elmer spectrum RX-1 IR spectrophotometer (Waltham, MA, USA).

X-ray Photoelectron Spectroscopy (XPS) measurements were carried out in a Specs GmbH system, equipped with a XR50 double anode source, a Specs Astraios 190 electron analyser with an angular acceptance of 60° and a micro channel plate 2D CMOS detector. The spectra were recorded using non-monochromatized Al K$_\alpha$ as primary excitation line. Sample EG1 was observed, upon functionalization with Ni NPs, in as-prepared conditions and after two annealing steps, the first at 350°C for 20 minutes, and the second at 550°C for 20 minutes. We point out that the particle analysis carried out on the AFM data reveals that the coverage of Ni NPs on the graphene is about 15%. Considering the geometry of the nanoparticles, the actual contribution from the Nickel atoms to the XPS signal is very limited. For this reason, the signal that we get for the Ni 2p core level, even after several hours, is barely sufficient for distinguishing its main contribution and this is the reason why we decided not to include more components (like the shake up satellite, for example) than those that were apparent, for analyzing the spectrum. The fitting of the core levels has been done by considering an iterative Shirley-type background and using symmetric Voigt lineshapes (or doublets) for every component, except for graphene, where we have used a Gaussian-broadened Doniach-Sunjic function with asymmetry factor 0.1.

Scanning tunneling microscopy (STM) and scanning tunneling spectroscopy (STS) measurements were performed in an ultra-high vacuum RHK Technology STM with a base pressure of $5 \cdot 10^{-11}$ mbar with tungsten tips electrochemically etched in a 2 M NaOH solution. The tips are degassed in situ, and subsequently, flashed by applying 600 V between tip (positive) and filament (negative) and then quickly increasing the filament current until 10 μA of emission current is detected. This procedure removes the oxide from the tips. The sample was degassed overnight at 200°C, and STM topographies were acquired after an additional annealing of the sample at 550°C for 30 minutes. The measurements were carried out in constant current mode with a tunneling current of 0.3-0.5 nA. Gwyddion software package was used to analyze STM images.

**Acknowledgements**

Authors want to thank Aldo Moscardini for his help in DLS analysis. This work has received funding from PNRR MUR project PE00000023 – NQSTI.

**Conflict of Interest**

The authors declare no conflict of interest.

# Supporting Information

**Rapid synthesis of uniformly small nickel nanoparticles for the surface functionalization of epitaxial graphene**

*Ylea Vlamidis*[1,2,*], *Stiven Forti*[3,4], *Antonio Rossi*[3,4], *Carmela Marinelli*[2], *Camilla Coletti*[3,4], *Stefan Heun*[1], *Stefano Veronesi*[1,*]

[1]NEST, Istituto Nanoscienze–CNR and Scuola Normale Superiore, Piazza San Silvestro 12, 56127 Pisa, Italy

[2]Department of Physical Science, Earth, and Environment, University of Siena, Via Roma 56, 53100 Siena, Italy

E-mail: stefano.veronesi@nano.cnr.it; ylea.vlamidis@unisi.it

[3]Center for Nanotechnology Innovation@NEST, Istituto Italiano di Tecnologia, Piazza San Silvestro 12, Pisa 56127, Italy

[4]Graphene Laboratories, Istituto Italiano di Tecnologia, Genova 16163, Italy

## 1. Raman analysis of mono/bilayer epitaxial graphene sample (EG1) and morphological characterization after thermal annealing

EG samples were grown on 6H-SiC as outlined in the Experimental section. Firstly, Raman spectroscopy was exploited to assess the quality and composition of EG1. The Raman spectrum of EG1 (**Figure S1a**) shows the typical graphene G and the 2D peaks, centered at 1597 cm$^{-1}$ and 2715 cm$^{-1}$, respectively. The full width at half maximum (FWHM) of the 2D band is between 30 and 50 cm$^{-1}$ (average value 37.8 cm$^{-1}$), and is located between 2715 cm$^{-1}$ and 2736 cm$^{-1}$, values which are characteristics of mono-bilayer graphene.[1] The intensity of the D peak, related to the presence of defects, is negligible, indicating the high quality of EG1. The peaks at 1520 and 1713 cm$^{-1}$ are attributed to the SiC substrate and represent the overtone of the TO phonon and the combination of optical phonons with wave vectors near the M point at the zone edges, respectively.[2]

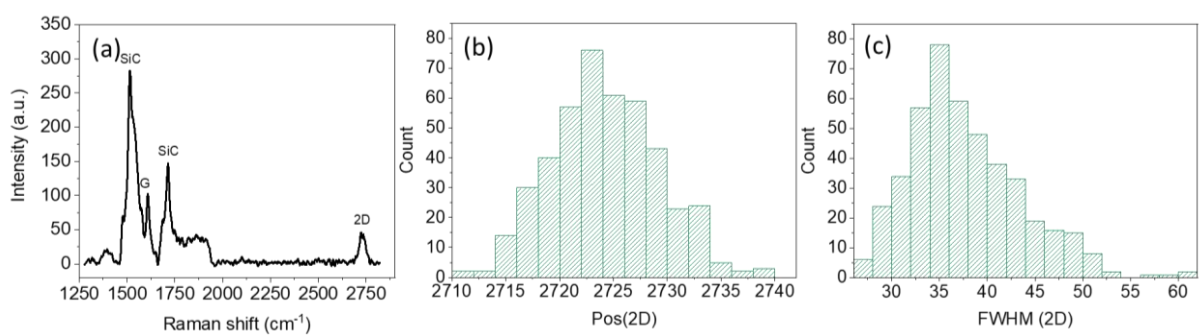

**Figure S1.** (a) Averaged Raman spectrum of EG1 (map area 21x21 μm²). Histograms of Raman parameters presenting (b) distribution of 2D band position and (c) FWHM.

## 2. Morphological characterization of EG1 sample after thermal annealing

AFM topographies were acquired for Ni NPs on EG1 after annealing the sample at 550°C. As shown in **Figure S2** the thermal process does not alter the surface morphology, *i.e.* cluster dimensions and distribution.

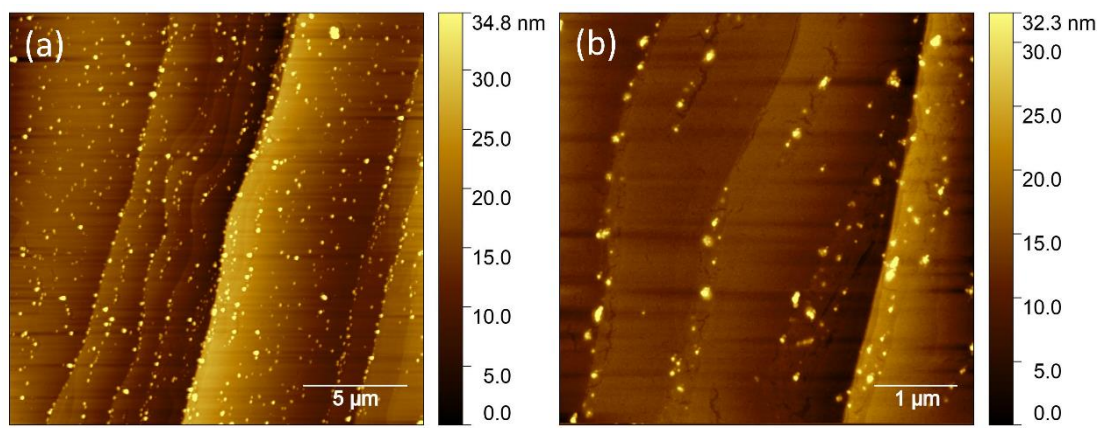

**Figure S2.** AFM topographies at different magnification of Ni NPs on EG1 after annealing the sample at 550°C.

## 3. Morphological and spectroscopic characterization of sample EG2

**Figure S3** shows the typical Raman spectra acquired both in buffer layer and monolayer areas (a), 2D intensity analysis (b), and the AFM height image (c) obtained for the EG2 sample employed for STM studies.

Combined Raman and AFM measurements show that the sample EG2 is constituted by buffer layer and monolayer graphene. Indeed, spatially resolved Raman spectra show large areas of buffer layer characterized by the typical SiC bands, and no graphene-related bands are detected (**Figure S3**, panels **a** and **b**). Nevertheless, in several areas the typical 2D band of monolayer graphene was detected, characterized by a single Lorentzian peak cantered at 2740 cm$^{-1}$ with an average full width at half maximum (FWHM) of 52 cm$^{-1}$, which suggests the presence of compressive strain.[3] Consistently, AFM images of the sample (**Figure S3c**) reveal large areas of buffer layer, with monolayer coverage of about 30%.

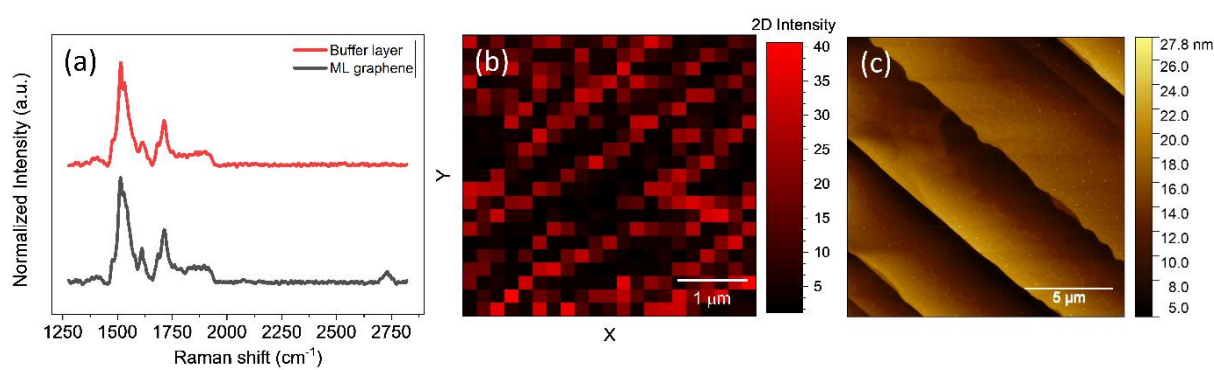

**Figure S3.** Characterization of sample EG2: (a) Raman spectra of monolayer graphene (ML) and buffer layer areas, (b) spatial Raman map (21 μm × 21 μm) of the 2D peak intensity, where black areas relate to buffer layer regions. (c) AFM height image of sample EG2.

## 4. Further Microscopy Data

After the functionalization with Ni NPs, differently from what was observed for the previous EG sample, selective pinning of Ni at graphene step-edges and coalescence are not observed. Instead, the dispersion of NPs on the surface of sample EG2 is uniform (refer to **Figure S4**), probably due to the higher roughness and corrugation of the surface, typical of the buffer layer as a consequence of the covalent bonds between the carbon atoms and the silicon atoms of the SiC(0001) surface.[4]

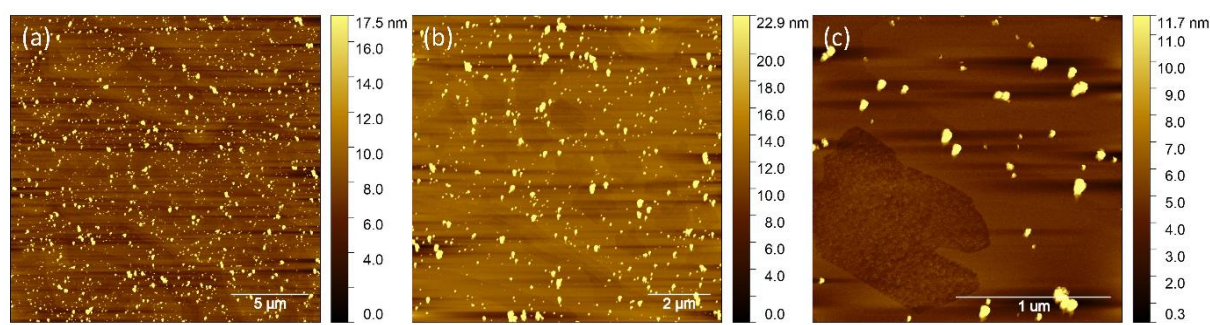

**Figure S4.** AFM images at different magnifications of the as-prepared Ni NPs on sample EG2 (buffer-monolayer).

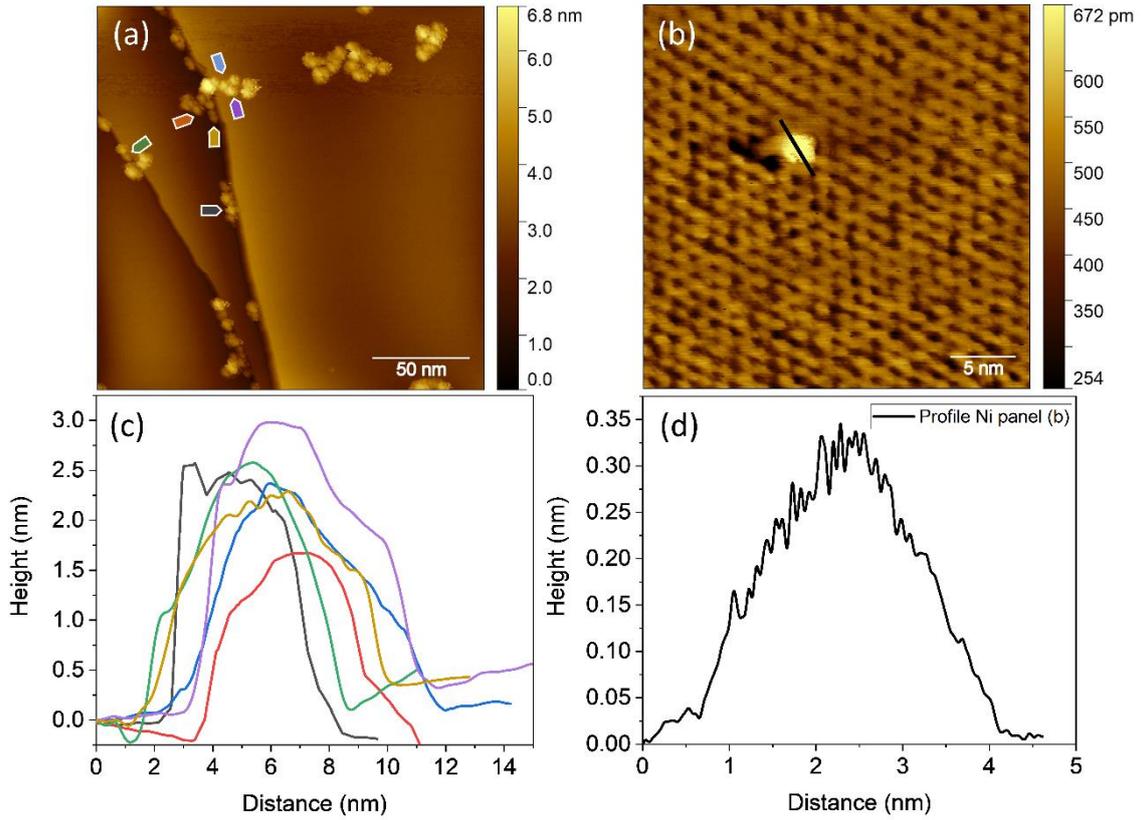

**Figure S5.** (a and b) STM images along with the respective line profiles (c and d) of Ni NPs on EG2. The considered NPs in panel (a) are highlighted by arrows, whose colors correspond to those of the height profiles in panel (c). The STM images are also reported in Figure 5 of the main text, panels b and e, respectively.

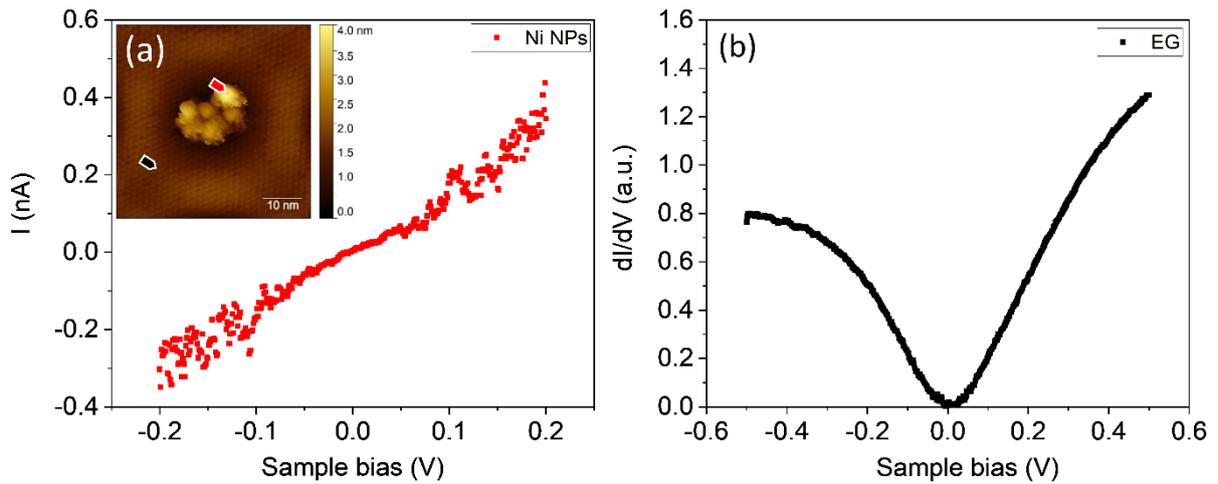

**Figure S6.** Representative STS measurements on Ni NPs on EG2: (a) I-V curve acquired on a cluster of Ni NPs, and (b) dI/dV spectrum recorded on the graphene layer. The curves were averaged over five measurements acquired at the positions marked by the arrows in the inset of (a). STM scan: 50x50 nm$^2$, $V_s = 0.50$ V, $I_t = 0.4$ nA.